\documentclass[graybox]{svmult}

\usepackage{tablefootnote}
\usepackage{type1cm}
\usepackage{float}
\usepackage{xurl}% activate if the above 3 fonts are
\usepackage{makeidx}         % allows index generation
\usepackage{graphicx}        % standard LaTeX graphics tool
\usepackage{multicol}        % used for the two-column index
\usepackage[bottom]{footmisc}% places footnotes at page bottom

\usepackage{newtxtext}       % 
\usepackage[varvw]{newtxmath}       % selects Times Roman as basic font
\usepackage[english]{babel}
\usepackage{url}

\date{}
\usepackage{amsmath}
\usepackage{tcolorbox}
\usepackage{graphicx}
\usepackage{comment}
\usepackage[colorlinks=true, allcolors=blue]{hyperref}
\usepackage{array}
\usepackage{blindtext}
\usepackage{tcolorbox}
\usepackage{graphicx}
\usepackage{xcolor}
\usepackage{nameref}
\usepackage{hyperref}
\usepackage{wrapfig}
\usepackage{float} % For improved floating control
\usepackage{placeins}
\usepackage{url}

\definecolor{lightgray}{gray}{0.95}

\title*{From misinformation to climate crisis: Navigating vulnerabilities in the cyber-physical-social systems}
\titlerunning{From misinformation to climate crisis}

\author{Tooba Aamir\orcidID{0000-0001-6190-1863} and\\ Marthie Grobler\orcidID{0000-0001-6933-0145} and\\
Giovanni Russello \orcidID{0000-0001-6987-0803}}
\authorrunning{Aamir et al.}

\institute{Tooba Aamir \at CSIRO's Data61, Australia, \email{tooba.aamir@data61.csiro.au}}

\begin{document}
\maketitle

\begin{abstract}
\\
Within the \textit{cyber-physical-social-climate nexus}, all systems are deeply interdependent: cyber infrastructure facilitates communication, data processing, and automation across physical systems (such as power grids and networks), while social infrastructure provides the human capital and societal norms necessary for the system's functionality. Any disruption within any of these components, whether due to human error or system mismanagement, can propagate throughout the network, amplifying vulnerabilities and creating a significantly scaled impact. 
This chapter explores the critical role of human vulnerabilities within the cyber-physical-social-climate nexus, focusing on the interdependencies across cyber, physical, and social systems and how these interdependencies can scale in a climate context. While cyber and physical vulnerabilities are readily apparent, social vulnerabilities (such as misinformation, resistance to policy change, and lack of public awareness) often go unaddressed despite their profound impact on resilience and climate adaptation. Social infrastructure, including human capital, societal norms, and policy frameworks, shapes community responses and underpins adaptive capacity, yet it is also a significant point of failure when overlooked. 
This chapter examines how human cognitive biases, risk misperception, and decision-making silos within interconnected systems can lead to resource misallocation and weakened policy effectiveness. These factors are analyzed to demonstrate how inadequate responses across cyber-physical-social layers can cascade, amplifying climate-related risks. By addressing these human factors and aligning decision-making frameworks, we aim to strengthen resilience and foster cohesive adaptation strategies that account for the intricate interrelations of cyber-physical-social-climate systems.
\end{abstract}


\section{Introduction}
The cyber-physical-social-climate nexus underscores the intricate interplay of each domain, each integral to modern society's functionality and resilience. These domains are deeply interconnected, with \textit{cyber} infrastructure enabling communication, automation, and data-driven decision-making across \textit{physical} systems, while \textit{social} infrastructure provides the human capital, societal norms, and regulatory frameworks that sustain their operation. The cyber-physical-social nexus underscores that the performance and resilience of any one system are heavily influenced by its interactions with the others \cite{haque2021realizing}. Adding \textit{climate}-related interactions to this already complex nexus brings about an array of new complications. 

As the world faces the escalating impacts of extreme weather events, the importance of understanding the cyber-physical-social-climate nexus becomes increasingly evident. Rising temperatures and other climate-related hazards threaten critical infrastructure, increasing the risk of cascading failures across interconnected systems. In addition, the role humans play is of increasing concern. Although climate and weather are distinct, with climate referring to long-term atmospheric patterns and weather to short-term conditions, climate risks are scientifically linked to shifts in the frequency and intensity of certain weather extremes. Advances in event attribution science have strengthened the connection between human-caused climate risks and specific extreme events, such as the 2020 Australian bushfires and European heatwaves \cite{ipcc2023}. For instance, misinformation about energy policies or climate adaptation measures can delay necessary upgrades to critical infrastructure or erode public trust in mitigation strategies \cite{lse2024}. These challenges are compounded by the deliberate spread of disinformation by vested interests aiming to undermine climate action. Studies show that such disinformation campaigns not only skew public perception but also hinder effective policy implementation by amplifying doubts about the scientific consensus on climate risks \cite{nature2023}. Addressing these vulnerabilities requires a holistic approach that safeguards individual systems and interdependencies to maintain essential societal functions like energy supply and public safety. 

Of particular interest is new risks tied to human factors, such as errors, cognitive biases, and decision-making silos, that are introduced with new technological advancements \cite{grobler2024building}. 
We specifically consider the escalating challenge posed by digitally disseminated climate misinformation and disinformation — a crucial vulnerability in today’s interconnected world \cite{cook2019understanding}. This information significantly skews public perception and obstructs effective policy implementation, posing widespread challenges that threaten the integrity and adaptability of interconnected systems. Such issues demand heightened attention from governments and international bodies, as they are central to safeguarding national and global security infrastructure against emerging threats. 
Addressing these vulnerabilities requires a multifaceted approach that targets enhanced digital literacy and situational awareness to improve the public’s ability to respond to crises with discernment. For instance, initiatives like the United Nations (UN)’ \textit{Verified} campaign aim to combat the `infodemic' of misinformation by disseminating reliable, fact-based content \cite{UNVerified2020}. Furthermore, improving cross-sectoral coordination and integration of cyber, physical, and social strategies is essential for developing resilience against the compounded risks posed by misinformation in the face of climate risks. 

Human involvement in the cyber-physical-social-climate nexus introduces several vulnerabilities, primarily due to limited situational awareness, prevalent cognitive biases, and the widespread amplification of misinformation. These vulnerabilities become particularly acute during natural disasters and crises, where cybercriminals can exploit them. For example, phishing attacks during such times often play on the urgency and fear surrounding an event, tricking the public into divulging sensitive information or donating to fraudulent causes\footnote{https://www.everbridge.com/blog/double-trouble-when-climate-change-and-cyber-crime-collide/} \cite{Zhuo24}. The aftermath of Hurricane Sandy saw a notable increase in such scams, with criminals taking advantage of the chaos and the outpouring of global support to launch various phishing schemes aimed at siphoning funds intended for disaster relief\footnote{https://www.forbes.com/sites/causeintegration/2015/10/05/charity-scams-put-the-disaster-in-disaster-relief/}. 
Another example was during the 2020 Australian bushfires, when misinformation regarding the cause of the fires spread rapidly through social media, misleading the public and placing additional strains on emergency response efforts. This lead to misplaced public anger and hinder the efficient allocation of resources, demonstrating how false information can exacerbate the real-time management of disaster responses \cite{inquiry2020final}. 

Similarly, during the \mbox{COVID-19} pandemic, a myriad of conspiracy theories about the virus origins and vaccine safety proliferated, significantly impacting public health responses and vaccine uptake rates worldwide \cite{islam2021covid, pertwee2022epidemic}. Sentiment analysis of over 2 million tweets showed a 35\% rise in negative sentiment linked to vaccine conspiracy hashtags (e.g.,~\#Plandemic, \#DoNotComply), resulting in a decline in public trust correlated with vaccination hesitancy in the US, UK, and Brazil \cite{islam2021covid, pertwee2022epidemic}.  Overall, recent studies indicate a significant decline in public trust towards institutions handling climate risk initiatives. This erosion hampers effective communication and policy implementation and a sentiment shifts after misinformation spikes (e.g., –25\% trust drop post-\#ArsonEmergency, +22\% sentiment gain in UN \textit{Verified} campaign) \cite{islam2021covid,UNVerified2020,edelman2023climate}. Therefore, social systems, including public perceptions, misinformation, and policy resistance, often serve as amplifiers of vulnerabilities across the nexus.

This chapter looks at the vulnerabilities that relate to human behavior in this cyber-physical-social-climate nexus, examining how human actions, decisions, and behaviors can compromise the integrity and adaptability of interconnected systems. 

It highlights how climate misinformation propagates through digital channels, distorts risk perception, fuels policy resistance, and compromises the effectiveness of critical infrastructure systems. Through empirical case studies and evidence-based analysis, the chapter develops an integrated set of recommendations that combines technological, policy, and community-based interventions to enhance system-wide resilience. By tackling the root causes of misinformation and enhancing the capacity of critical infrastructures to withstand and adapt to these challenges, we can significantly improve the resilience of societies against the cascading effects of misinformation in the cyber-physical-social-climate nexus. The primary objectives of this chapter are to:
\begin{itemize}
    \item Examine how climate misinformation and disinformation affect interdependent cyber, physical, and social systems, contributing to cascading risks and system-wide disruptions (Sections 2 and 3).
    \item Assess the amplification of infrastructure vulnerabilities during natural disasters and emergencies due to communication failures and misleading information (Section 3).
    \item Evaluate current and emerging strategies, including governance initiatives, cybersecurity measures, and artificial intelligence (AI) tools, to counter misinformation and protect information integrity (Sections 4 and 6).
    \item Explore actionable, cross-sectoral interventions to enhance public trust, bolster infrastructure resilience, and improve climate adaptation outcomes across the CPS-Climate nexus (Sections 5 and 7).
\end{itemize}

By addressing these objectives, the chapter provides a comprehensive overview of the multifaceted challenges presented by misinformation within the cyber-physical-social-climate nexus, offering insights into how these challenges can be effectively managed and mitigated through coordinated efforts across multiple domains.

\section{Misinformation and disinformation: Definitions, differences, and systemic impacts}

Within the cyber-physical-social-climate nexus, social vulnerabilities represent a significant challenge that, when unaddressed, can amplify risks and create substantial barriers to effective adaptation and response. Social systems, encompassing societal norms, human capital, and policy frameworks, are crucial in supporting the functionality of both cyber and physical systems and enabling resilience. 

Although misinformation and disinformation are both forms of false information, they differ in intent. \textit{Misinformation} is shared unintentionally, while disinformation is deliberately constructed and disseminated with the intention of misleading. It often arises from misunderstandings or errors and spreads through digital platforms and social networks. Misinformation fosters confusion, scepticism, and resistance among stakeholders, which can hinder the effective operation of cyber and physical systems. This includes resistance to adopting secure systems or implementing critical updates, increasing vulnerability to cyberattacks~\cite{schreiber2021defalsif}. \textit{Disinformation}, on the other hand, is deliberately created and disseminated with the intent to deceive and manipulate public opinion or disrupt systems. It is a crafted falsehood designed to manipulate individuals or groups and often involves organised campaigns to destabilise trust in factual information and institutions \cite{hameleers2023disinfo}. This distinction is critical in framing our analysis. While disinformation is a growing threat, this chapter primarily focuses on misinformation and its cascading effects on trust, policy resistance, and system resilience, acknowledging disinformation where relevant.

\subsection{Systemic breakdowns: Impact on the cyber-physical-social nexus}
Climate mis- and disinformation have a major impact on the cyber-physical-social nexus. It weakens the effectiveness and resilience of these interconnected systems and erodes societal trust in climate adaptation strategies, further complicating the challenges faced by these integrated networks. 

In the \textit{cyber} domain, misinformation about the security of system upgrades or new technologies can cause delays in adopting essential cybersecurity measures\footnote{https://www.weforum.org/stories/2021/10/the-next-decade-will-be-defined-by-climate-change-and-cyber-risks/}. Such hesitation exposes cyber systems to increased risks of attacks, especially during critical periods such as natural disasters or significant public events, when robust defences are most needed. Cybersecurity frameworks such as the EU’s Digital Services Act\footnote{https://commission.europa.eu/strategy-and-policy/priorities-2019-2024/europe-fit-digital-age/digital-services-act} and platforms such as Meta’s SimSearchNet++\footnote{https://ai.meta.com/blog/heres-how-were-using-ai-to-help-detect-misinformation/} exemplify proactive technical measures that identify, mitigate, and prevent the amplification of false content, including through bot detection, content provenance tools, and systemic risk audits.  Instances of misinformation undermine the overall security posture of cyber infrastructure, increasing its vulnerability to cyberattacks. The impact extends into the \textit{physical} domain, where misinformation about climate adaptation measures can lead to resistance. This resistance can result in project delays or inadequate implementations related to upgrading flood defences or installing renewable energy systems, which leaves communities more vulnerable to the exacerbated effects of climate risks, such as increased frequency and severity of extreme weather events. The consequences of misinformation are most profound in the \textit{social} domain. Misinformation erodes public trust in scientific data and governmental agencies, which is crucial for collective climate action. It polarises public opinion and creates significant barriers to achieving a consensus for implementing large-scale climate resilience and adaptation strategies. False narratives, while sometimes intentionally created, often occupy a grey area between misinformation and disinformation. They may blend unintentional errors with deliberate manipulation, making their classification complex. This is particularly evident in contexts such as the COVID-19 pandemic, where both misinformed sharing and targeted disinformation coexisted \cite{islam2021covid,pertwee2022epidemic}. False narratives are particularly destructive in the climate context. Coordinated false narratives campaigns can drastically undermine public support for environmental policies by fostering scepticism about the reality of climate risks.

Beyond its immediate impacts on public perception and infrastructure readiness, climate misinformation also undermines broader systemic responses to climate risks. It contributes to policy resistance by fueling scepticism and division among stakeholders, delaying or obstructing necessary regulatory measures. Furthermore, by undermining scientific consensus, misinformation erodes trust in climate science and institutions, reducing the credibility of evidence-based adaptation and mitigation policies. It also distracts from viable climate solutions, redirecting attention and resources toward debunked theories or unproven alternatives. At an international level, misinformation disrupts global cooperation by fostering distrust among nations and complicating collective climate action. These effects illustrate how misinformation not only compromises public understanding but also weakens the institutional and diplomatic frameworks essential for coordinated climate resilience.

In the physical domain, \textit{short-term effects} during emergencies such as natural disasters, misinformation can have immediate harmful effects \cite{maertens2020combatting}. For example, false information about evacuation routes, resource availability, or emergency protocols can delay responses, create panic, and worsen cascading failures in cyber and physical systems. A notable instance occurred during the 2022 floods in Lagos, Nigeria, where the government had to combat false narratives by leveraging social media and reputable media outlets to provide live updates and factual information, hold press conferences, release official statements, and engage with community leaders to ensure accurate information reached a broader audience~\cite{GCoM2024}. 
Similarly, \textit{in the long run}, misinformation about physical infrastructure projects, such as automated transportation systems or smart cities, can result in delays, resistance, or even rejection of essential upgrades, leaving physical systems susceptible to disruptions ~\cite{CISC2024RR}. An example of this is the reluctance to adopt innovative grid technologies, making energy infrastructure vulnerable to inefficiencies and failures. In addition, distrust in cyber technologies, fuelled by misinformation, amplifies vulnerabilities across multiple domains, impeding the adoption of advanced cybersecurity measures and innovative infrastructure solutions.

From a social perspective, the effects of misinformation also unfold across temporal dimensions. \textit{In the short term}, i.e., the response phase, spreading false information during emergencies, such as incorrect evacuation routes, erroneous safety guidelines, or fabricated resource shortages, can cause immediate confusion, panic, and misallocation of critical resources. For example, during the 2022 floods in Lagos, Nigeria, inaccurate information about evacuation zones delayed resident responses and diverted emergency services \cite{GCoM2024}. Evacuation was delayed between three and six hours in certain districts due to the circulation of false information about flood zones on WhatsApp and Facebook. Similarly, during the 2011 Fukushima Daiichi nuclear disaster in Japan, false claims about radioactive contamination triggered mass panic buying, resulting in shortages that impaired supply chains \cite{owen2019nuclear}. 
In contrast, \textit{long-term effects} are often subtler but more corrosive. Misinformation that undermines trust in climate science or technologies (e.g., renewables) can result in sustained public scepticism, resistance to adaptation policies, and hesitancy in adopting resilience-enhancing infrastructure\footnote{https://www.theguardian.com/environment/2025/jun/19/climate-misinformation-turning-crisis-into-catastrophe-ipie-report}. The 2024 Global Risks Report ranks misinformation and disinformation as the most significant short-term risk to human society, with extreme weather events as the top long-term risk, highlighting the critical impact of obscuring facts about climate risks~\cite{WEF2024}.  

Misinformation can also cause cascading failures across the cyber-physical-social nexus and exacerbate vulnerabilities in critical infrastructure, such as power grids, water systems, and healthcare facilities, particularly when these infrastructures are already under stress due to climate impacts. For instance, false information about an impending natural disaster can lead to unnecessary panic, resulting in premature and uncoordinated evacuations that strain resources and may hinder genuine emergency response efforts. Misinformation about resource availability or emergency procedures during climate hazards can also lead to mismanagement of resources, further destabilizing already vulnerable systems and potentially leading to catastrophic failures. Similarly, campaigns that target climate risks data can lead to scepticism about the severity of climate threats, reducing public support for essential policy measures and investments in resilience-enhancing infrastructure.

\subsection{Misinformation and policy resistance}

Misinformation fuels policy resistance when stakeholders (regardless of whether they are individuals, organizations, or governments) oppose regulatory changes aimed at enhancing resilience \cite{ecker2022psychological,kozyreva2023incorporating}. The large-scale cross-country survey by International Monetary Fund (IMF) in 2023 shows that public support for climate policies increases by up to 7\% when co-benefits are communicated, while cost-emphasizing messages reduce support by 9\% \cite{dablanorris2023public}. This resistance stems from economic concerns, political ideologies, distrust in government or technology, and reluctance to shift from established practices. 

Misinformation about the implications of new policies or technologies exacerbates this resistance, leading to delays or the complete blocking of necessary upgrades critical to ensuring the resilience of cyber and physical infrastructure \cite{ecker2022psychological}. For example, in 2022–24, fossil-fuel interests used denial and greenwashing campaigns to delay EU climate regulations and block renewable energy policies. These disinformation efforts directly slowed policy rollouts\footnote{https://www.theguardian.com/environment/2025/jun/19/climate-misinformation-turning-crisis-into-catastrophe-ipie-report}. Similarly, fake narratives that misattribute disaster causes (e.g., blaming High-frequency Active Auroral Research Program (HAARP) for hurricanes) complicate attempts to implement evidence-based disaster policies, hindering resilience planning and systematic reforms\footnote{https://disa.org/bloomberg-discusses-disaster-misinformation-with-uc-professor/}. Such policy resistance delays the implementation of necessary upgrades, thereby increasing vulnerability to future disasters. Over time, misinformation embedded within public discourse may institutionalise inaction or erode political will. 

The spread of misinformation contributes to gaps in community awareness and trust, which are crucial for addressing vulnerabilities within the nexus~\cite{OECD2022}. Many individuals and organizations lack a comprehensive understanding of the risks associated with weak cybersecurity hygiene or the importance of proactive climate adaptation, leading to poor cybersecurity practices and inadequate preparation for extreme weather events. A pertinent example is the opposition to wind farm developments in New South Wales, Australia, where false claims about adverse health effects, environmental degradation, and economic inefficiency delayed several major projects. Misinformation disseminated via online forums contributed to local councils withdrawing support, illustrating how disinformation can stall renewable energy transitions \cite{oilprice2024misinformation, longview2014wind}.

Enhanced community awareness and strengthened policy frameworks can counter the detrimental effects of misinformation and disinformation \cite{longview2014wind,roozenbeek2023countering}. Awareness and open discussions for community involvement are also pivotal; increasing digital literacy and raising awareness about climate adaptation prepare stakeholders to recognize and reject misinformation. Building a culture of trust and transparency in government and technology adoption can help mitigate resistance to necessary changes. Engaging communities in the policymaking process and providing clear, factual information about the benefits and challenges of new policies and technologies fosters a supportive environment for systemic resilience \cite{Trijsburg2024Disinformation}. By addressing the misinformation that often underlies policy resistance and public unawareness, stakeholders across the cyber-physical-social-climate nexus can enhance their collective ability to withstand and adapt to emerging challenges, ensuring a more resilient infrastructure and a better-prepared society.

\section{Impacts of misinformation on critical infrastructure}
Misinformation has become a significant threat to critical infrastructure, particularly during climate hazards and natural disasters. Specifically due to the broad scope of interconnected infrastructure sectors and the wide reach of associated essential services, any inaccurate information shared could have far-reaching impacts, particularly if the users do not take the time to verify and validate the information before sharing \cite{jamalzadeh2022protecting}. Not only does it undermine public trust and disrupt emergency responses, but it also amplifies vulnerabilities in essential systems such as power grids, water supplies, and healthcare, and unduly influences commodity consumption. When these essential systems are involved, people can very easily become riled up, causing false and misleading information to spread even further and faster. This can lead to increasing misperceptions and knowledge resistance \cite{Broda2024Misinformation}, which pose significant threats due to increased commodity consumption and resultant cascading failures.

\subsection{Misinformation during natural disasters}
Rapid information dissemination during natural disasters is vital for managing public response and ensuring safety. While social media can spread lifesaving information, it is also a source of misinformation \cite{Hilberts2024TheIO}. 
Misinformation during natural disasters has repeatedly demonstrated its potential to cause significant disruptions and harm, despite users believing that they are helping by sharing the information. Misinformation during a time in which people are particularly vulnerable and susceptible to external influences could cause public panic, hinder effective communication, and exacerbate the challenges faced by emergency responders \cite{Hilberts2024TheIO}. 
Key global incidents provide stark examples of the chaos and confusion that can ensue when false information is disseminated. 
These examples underscore the critical need for accurate information during natural disasters and highlight the detrimental effects misinformation can have on disaster management and public safety. 

A notable instance took place during the 2025 wildfires in Los Angeles County, United States when a software glitch resulted in an evacuation alert being sent to nearly 10 million residents, far beyond the actual danger zone. Compounding the error, outdated alerts were rebroadcast when cell towers resumed service after being offline, leading to widespread panic, traffic congestion and public confusion, delaying real evacuation orders to those in need \cite{downtoearth2025falsealerts}. The chaos included injuries from rushed evacuations and clogged routes that significantly delayed emergency services from reaching those in actual need\footnote{https://smdp.com/news/after-false-alerts-panic-millions-officials-move-to-state-system-for-evacuation-notification/}. Although system communication failures, such as software glitches or erroneous alerts, are not intended to mislead and thus fall outside the strict definition of misinformation, they can still erode public trust in emergency systems. This erosion may increase susceptibility to actual misinformation, as individuals become desensitized and in the long-run may choose to ignore future legitimate alerts.

Another example occurred in 2024 when misinformation complicated emergency responses during Hurricanes Helene and Milton. Conspiracy theories circulated, falsely claiming the hurricanes were government-engineered or that aid was being selectively withheld from certain political groups. These allegations threatened the safety of federal aid workers, delayed relief efforts and exacerbated tensions in communities already suffering from the hurricanes' impacts\footnote{https://www.bbc.com/news/articles/cx2lyzw7xwxo}. 
Another instance of harmful misinformation was during the 2019/20 bushfires in Australia (refer to Case Study~1), where the \textit{\#ArsonEmergency} hashtag inaccurately attributed the widespread devastation to arson, diverting attention from the emergency response and public safety. In this instance, the authority's ability to manage the crisis was undermined because resources were misallocated based on incorrect assumptions. This disinformation severely disrupted emergency coordination efforts and contributed to a significant erosion of public trust in climate policies \cite{bushfire_disinformation}. 

\subsection{Vulnerabilities amplified by misinformation}

Misinformation can significantly worsen vulnerabilities within critical infrastructure sectors, leading to delayed or inadequate responses to emerging challenges and crises. False information can cause confusion and disrupt critical operations due to unnecessary shutdowns or system overloads, or misinformation can be used specifically to exploit vulnerabilities, for example, a denial of service can occur after false information indicates a system vulnerability on a critical service.  

Several examples demonstrate how misinformation can impact various critical infrastructures, underscoring the importance of accurate information for effective management and resilience. Studies such as the 2023 Edelman Trust Barometer~\cite{edelman2023} show a significant decline in institutional trust following major misinformation events. Likewise, the Notre Dame Global Adaptation Initiative (ND-GAIN) country Index\footnote{\label{fn:ndgain}https://gain.nd.edu/our-work/country-index/rankings/} provides a benchmark for assessing climate adaptation readiness, which can be correlated with resilience to misinformation-induced delays.
In Africa, misinformation about climate-related healthcare risks, such as those linked to malaria and malnutrition, undermines efforts to implement effective adaptive measures during extreme weather events. Public confusion about the causes and preventive strategies for these health crises significantly delays effective responses, compromizing public health outcomes during critical periods \cite{Heffernan_2024}. Misinformation also disrupts water management, as false narratives about water scarcity can lead to ineffective or misguided strategies during droughts and floods. In vulnerable regions, conflicting information about water usage policies causes delays in the implementation of sustainable water management solutions, exacerbating the effects of water scarcity on communities and ecosystems \cite{Heffernan_2024}. 

In Taiwan, the energy sector faced challenges from misinformation campaigns that fuel distrust in renewable energy projects and the reliability of power grids. Misleading claims exaggerating the limitations of solar power or promoting nuclear energy as a more viable alternative have undermined public confidence in energy transitions necessary for achieving climate resilience \cite{liu2025climate}.  
During the 2019 European heatwave, disinformation campaigns exaggerated the inefficacy of renewable energy sources, claiming they were failing to provide the necessary power, stirring public and political unrest (refer to Case Study 2) \cite{strauss2024reporting, casero2023european}.
Furthermore, cybersecurity risks are increasingly exacerbated by misinformation campaigns that target critical infrastructure systems, exploiting vulnerabilities in interconnected networks such as power grids and transportation systems. These misinformation campaigns can significantly delay responses to real threats or amplify damage during cybersecurity crises, affecting the security and efficiency of critical operational processes \cite{CISC2024RR}. Consider the 2017 WannaCry ransomware attack, where misinformation fuelled fears, amplifying the impact of the attack and negatively impacting productivity because users were hesitant to switch on their computers for fear of being ransomed\footnote{https://www.cbsnews.com/news/wannacry-ransomware-attacks-wannacry-virus-losses/}.  

%%%%%%%%%%%%%%%%%%%%%%CASESTUDY 1%%%%%%%%%%%%%%%%%%%%%%%%%%%
%\begin{wrapfigure}{r}{1\textwidth}
%, float*=!ht
\begin{tcolorbox}[width=\textwidth,colback={lightgray},title={\textbf{Case Study 1: 2019/2020 Australian bushfires}},colbacktitle=gray,coltitle=white, label={box:info}]  
\fontsize{8pt}{10pt}\selectfont

The 2019/2020 Australian bushfire season \cite{inquiry2020final}, or \textit{Black Summer}, was one of the most devastating bushfire events in Australian history, burning over 18 million hectares, destroying thousands of homes, and causing significant loss of life and biodiversity. As the crisis unfolded, misinformation spread rapidly across digital platforms, with one of the most damaging falsehoods claiming that arson, rather than climate risks and hazards, was the primary cause of the fires. This misleading narrative was amplified by social media, some political groups, and climate sceptics, undermining scientific consensus that linked the severity of the fires to extreme climate conditions, including prolonged droughts and rising temperatures. 

 \begin{center}
        \includegraphics[width=1\linewidth]{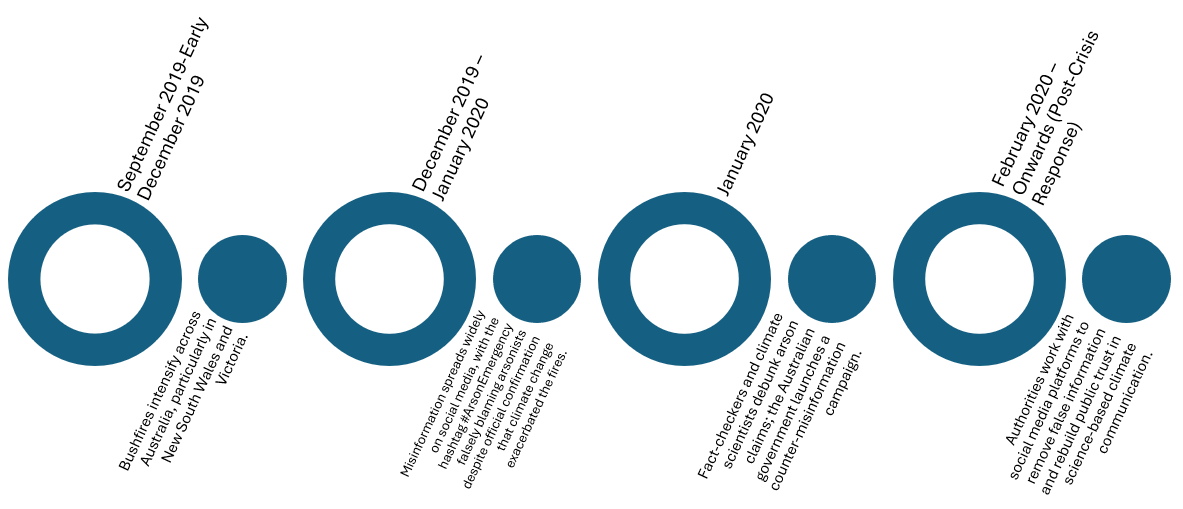} % Change "example.png" to your image filename
    \end{center}
    
Misinformation had widespread effects across the cyber-physical-social-climate nexus, influencing public perception, emergency response efforts, and climate policy, e.g., adaptation measures such as prescribed burning practices, and early warning systems, rather than emissions mitigation strategies.
  The effects included:\\
\\
\textbf{Cyber dimension: Digital networks and the spread of misinformation}
\begin{itemize}
    \item \textbf{Social media disinformation} (\textit{\#ArsonEmergency}) falsely shifted blame from climate risk to arson, amplified by bot networks and foreign actors, fueling polarization and distrust in climate science.
    \item \textbf{Emergency response hindered} as authorities diverted resources to counter misinformation, while fake donation scams misled donors, diverting aid from real relief efforts.
\end{itemize}

\textbf{Physical dimension: Infrastructure and environmental impacts}
\begin{itemize}
    \item \textbf{Misallocation of emergency resources}, delaying real responses. Distrust in evacuation orders caused confusion — some ignored warnings, others evacuated unnecessarily.
    \item \textbf{Weakened climate adaptation}, as misinformation reduced support for controlled burns and land reforms, contributing to long-term vulnerabilities.
\end{itemize}

\textbf{Social dimension: Public perception, policy resistance, and trust}
\begin{itemize}
    \item \textbf{Erosion of trust in authorities} complicated crisis management, as misinformation fuelled climate scepticism and delayed adaptation policies.
    \item \textbf{Misinformation weaponized by interest groups}, particularly fossil fuel lobbyists, to downplay climate risks and stall policy reforms.
\end{itemize}

During the Australian bushfires, misinformation disrupted crisis response and delayed climate action by eroding trust in science and weakening support for preventive measures like controlled burns. This underscores the need for stronger public trust, clear communication, and proactive misinformation management.\\
\\
\textbf{Key takeaways}
\begin{itemize}
    \item \textbf{Misinformation strains crisis response} by diverting resources and increasing confusion.
    \item \textbf{Digital platforms amplify misinformation}, spreading both facts and falsehoods.
    \item \textbf{Public trust is vital} as misinformation undermines science, policy, and emergency response.
    \item \textbf{Delays in climate adaptation and policy reforms} result from persistent misinformation.
    \item \textbf{Proactive misinformation management is essential}, requiring coordinated efforts across governments, media, and scientific institutions.
\end{itemize}

\end{tcolorbox} 
%\end{wrapfigure}

%%%%%%%%%%%%%%%%%%%%%%%%%%%%%%%%%%%%%%%%%%%%%%%%%%%%%%%%%%%

%%%%%%%%%%%%%%%%%%%%%%CASESTUDY 2%%%%%%%%%%%%%%%%%%%%%%%%%%%

\begin{tcolorbox}[width=\textwidth,colback={lightgray},title={\textbf{Case Study 2: 2019 European heatwave}},colbacktitle=gray,coltitle=white, label={box:info}]  
\fontsize{8pt}{10pt}\selectfont

In 2019, Europe experienced one of its most intense heatwaves, with temperatures reaching record highs in  countries such as France, Germany, and Spain \cite{strauss2024reporting, casero2023european}. This extreme weather event put immense pressure on energy grids, increased health risks, and disrupted daily life. Disinformation campaigns emerged, falsely claiming that renewable energy sources were failing and incapable of meeting energy demands. As the heatwave intensified, disinformation campaigns spread through social media and certain media outlets, arguing that solar and wind energy were unreliable and responsible for power shortages. This was despite evidence that renewable energy sources continued to perform effectively, and the actual strain on the grid was caused by increased electricity demand for air conditioning and cooling systems.

 \begin{center}
        \includegraphics[width=1\linewidth]{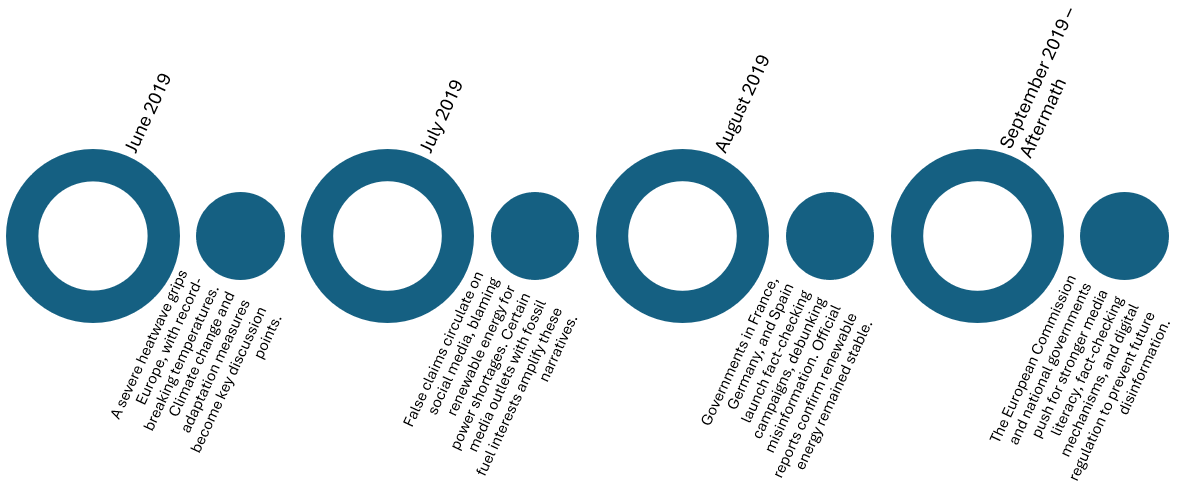} % Change "example.png" to your image filename
    \end{center}
    
Misinformation had widespread effects across the cyber-physical-social-climate nexus, contributing to public distrust in renewable energy, exacerbated political divisions, and influenced policy discussions on energy security. The effects included:\\
\\

\textbf{Cyber dimension: Digital networks and the spread of misinformation}
\begin{itemize}
    \item \textbf{Social media disinformation:} False claims on Facebook, Twitter, and YouTube blamed renewable energy for power shortages. Fossil fuel interest groups and climate sceptics spread doubts about solar and wind reliability.
    \item \textbf{Energy communication disruptions:} False reports about electricity prioritization caused panic and erratic energy use.
\end{itemize}

\textbf{Physical dimension: Infrastructure and energy system impacts}
\begin{itemize}
    \item \textbf{Grid strain:} Misleading claims led to increased energy use and overloading power networks.
    \item \textbf{Weakened climate adaptation:} Disinformation shifted blame from climate conditions to renewables, reducing support for sustainability, fueling policy debates and stalling green energy infrastructure.
\end{itemize}

\textbf{Social dimension: Public perception, policy resistance, and trust}
\begin{itemize}
    \item \textbf{Trust erosion:} False narratives framed renewable energy policies as failures, lowering confidence in climate strategies. 
    \item \textbf{Media distraction:} Fact-checkers focused on debunking false claims, delaying accurate crisis updates.
\end{itemize}

The case highlights cyber-enabled disinformation's role in undermining public confidence in climate-related infrastructure. The social dimension was particularly affected, as public mistrust fuelled political opposition to renewable energy projects. These disinformation narratives threatened to delay investment in critical energy transition policies, showing how false narratives can have long-term consequences on climate adaptation.\\
\\
\textbf{Key takeaways}
\begin{itemize}
    \item \textbf{Disinformation can shape political and economic decisions}, particularly in climate-related sectors.
    \item Social media plays a critical role in amplifying climate falsehoods, \textbf{requiring proactive counter-misinformation strategies}.
    \item \textbf{Government-led fact-checking initiatives} are essential in maintaining public trust in renewable energy transitions.
    \item \textbf{Improving media literacy} can help individuals critically assess false claims related to climate adaptation efforts.
\end{itemize}

\end{tcolorbox}

%%%%%%%%%%%%%%%%%%%%%%%%%%%%%%%%%%%%%%%%%%%%%%%%%%%%%%%%%%%

\subsection{Increased commodity consumption and cascading failure}
Misinformation can fuel behaviour that has unintended and often unforeseen consequences. 
Electric power utilities are a prime target of these commodity-driven misinformation campaigns due to their demand-response programs to reduce or shift electricity usage during peak periods in response to time-based rates or other financial incentives. 
Consider a scenario where a deceptive social media post is shared by a realistic looking fraudulent account. The message might incentivize increased electricity usage during peak times, citing a customer loyalty bonus. As the message spreads and more people attempt to specific electricity consumption, the likelihood of blackouts increases due to system overloads. This can lead to widespread impacts on public health and safety. If enough users in a specific region act on this misinformation, it can disrupt the power network \cite{jamalzadeh2022protecting}. Due to the recognized interdependency of most other critical infrastructure sectors on the energy sector \cite{kaur2024reag}, cascading infrastructure failures would follow with these interconnected  systems particularly vulnerable to misinformation-induced disruptions. For example, false narratives about power grid reliability can lead to public panic and overloading of energy systems during extreme weather events.

A real-world example of misinformation leading to increased commodity consumption and resulting in cascading failures across critical infrastructure is the 2011 Fukushima, Japan disaster. Following the earthquake and tsunami, misinformation and rumours about radiation levels and food and water safety spread rapidly. This led to panic buying and hoarding of essential commodities like bottled water, food, and fuel. 
The increased demand led to widespread supply shortages and strained supply chains, leading to cascading failures in the distribution of these critical resources. The misinformation amplified the impact of the disaster, making it more challenging for authorities to manage the situation and ensure the availability of essential supplies to those in need \cite{owen2019nuclear}.
% \vspace{-0.2em}

\section{Government strategies to counter misinformation}

Governments' responses are critical in combating misinformation that undermines public safety and security, particularly about critical infrastructure. Effective policy-making and governance are essential in addressing the challenges of misinformation, especially as they pertain to public awareness and resistance to necessary changes. 
The gaps in public awareness and widespread policy resistance pose significant obstacles to the resilience of the cyber-physical-social-climate nexus. These challenges are often exacerbated by misinformation that leads to public mistrust and hinders the adoption of necessary cybersecurity practices and climate adaptation strategies. In response, governments and policy-makers are tasked with devizing strategies that not only counter misinformation but also foster a more informed public and a more responsive infrastructural system.

Efforts to combat climate misinformation are gaining significant momentum globally as governments and intergovernmental organizations recognize accurate information's profound impact on societal resilience and critical infrastructure planning. To effectively combat the pervasive influence of misinformation, governments are deploying a range of strategic responses that aim to enhance public safety and bolster infrastructure resilience. These strategies include strengthening public awareness, increasing transparency, fostering collaborative governance, and leveraging technology and data \cite{Critical5_2024}. 

\subsection{Strategic awareness and governance initiatives}
Enhancing public awareness and fostering collaborative governance is pivotal. By investing in comprehensive awareness initiatives, governments can significantly improve public understanding of cybersecurity risks and the urgency of climate adaptation. Globally, several initiatives exemplify this approach.

For example, initiatives like the UN's \textit{Verified} campaign, launched in May 2020, aimed to combat COVID-19 misinformation by disseminating accurate, science-based information \cite{UNVerified2020}. Its primary mission was to disseminate credible, science-based content on science, solidarity, and solutions. It further supported the “digital first responders” to amplify UN-verified content across social media and messaging platforms. The campaign achieved substantial reach and engagement in its first three years, partnering with 250 organizations, collaborating with 200 influencers, and generating 1.4 billion engagements across social media platforms. It reportedly reached over 1 billion people globally by 2023. Panel testing of the \textit{Verified Champions}’ content in 2022 showed significant increases in knowledge levels and intentions to act on climate misinformation\footnote{https://www.purpose.com/case-study/verified/}. While explicit sentiment metrics reported by the UN are limited, insights can be drawn from the WHO’s broader “infodemic management” frameworks, which highlight sentiment change as a key evaluation factor. The campaign also successfully recruited millions of volunteer “digital first responders”, encompassing grassroots to national-level networks, including initiatives led by country-specific hubs.

The \textit{Global Initiative for Information Integrity on Climate Change}, launched by UNESCO, the UN and Brazil, aims to strengthen global information integrity and combat climate-related mis- and disinformation  \cite{UNESCO2024}. This initiative includes a Global Fund dedicated to research on disinformation and supports campaigns designed to safeguard accurate climate information. Specific goals include exposing disinformation, strengthening global cooperation, and enhancing public communication campaigns. Similarly, the European Commission's \textit{Communication on Managing Climate Risks} commits to combating climate disinformation by addressing its impacts on public trust, policy implementation, and international cooperation \cite{EC2024}. Australia’s \textit{National Climate Resilience and Adaptation Strategy} (2021–2025) emphasizes collaboration to improve public understanding of climate risks and counter misinformation as part of broader adaptation efforts \cite{AustraliaC2021}. These initiatives focus on developing accessible and engaging programs that provide clear, accurate information about the threats posed by misinformation. Such efforts are crucial for empowering citizens to critically evaluate information and make informed decisions. 
UNESCO's \textit{Global Initiative for Information Integrity on Climate Change} seeks to address climate-related disinformation by supporting research and communication campaigns. The initiative aims to produce evidence on the impact of disinformation and supports policies that protect information integrity \cite{unesco2024information}. 

Collaborative governance plays a crucial role in strengthening these efforts. This involves a multi-sectoral approach where governments collaborate with international bodies, private sectors, and non-governmental organizations \cite{Critical5_2024}. Building public trust through transparent governance and regular, clear communication helps to demystify policy decisions and infrastructure developments. Governments providing regular updates and engaging directly with citizens through open forums and discussions help reduce the opportunities for misinformation to take root. The \textit{Global Covenant of Mayors for Climate and Energy} exemplifies how cities can combat disinformation through transparency, community engagement, and partnerships with media outlets to debunk false narratives. Examples include Cape Town’s real-time updates on water levels and Nairobi's transparent communication during development projects \cite{GCoM2024} and the G20 Summit Initiative, announced in 2024, involving countries like the UK, France, Chile, Morocco, Denmark, and Sweden working to address climate disinformation through shared strategies and resources \cite{G202024}. The World Meteorological Organization also contributes by disseminating scientifically rigorous information and developing tools to enhance public understanding of climate science \cite{WMO2024}. Intergovernmental security-focused programs such as the NATO Science for Peace and Security (SPS) Programme\footnote{\url{https://www.nato.int/cps/en/natohq/78209.htm}} also present strategic opportunities to strengthen information integrity and societal resilience against climate-related misinformation.

\subsection{Technology and data integration}
Leveraging technology and data is crucial in the fight against misinformation. Advanced technologies, such as data analytics and machine learning, enable governments to monitor trends in misinformation, identify sources of false information, and develop precise responses that neutralize misleading narratives before they spread widely. These technological tools are indispensable for swiftly countering misinformation in an increasingly connected world. 

Multiple nations and organizations have called for holding social media platforms accountable for spreading climate misinformation, urging the adoption of universal definitions of climate disinformation and promoting information integrity~\cite{CAADOpenLetter2024}. Consider the monitoring systems at COP Conferences that actively track and counteract misinformation during global climate summits to ensure accurate narratives dominate discussions \cite{CAAD2024}.
Complementing these efforts, projects like \textit{defalsif-AI} are developing advanced media-forensic tools to evaluate the credibility of online content. Focused on detecting politically motivated disinformation, \textit{defalsif-AI} aims to safeguard democratic processes and public trust~\cite{schreiber2021defalsif}.  These tools utilize audio-visual media forensics, text analysis, and the fusion of these technologies with AI to identify and assess the authenticity of digital content.

To reinforce technological resilience against misinformation, a robust suite of cybersecurity tools is emerging to address the full spectrum of mis- and disinformation threats. These tools span legal, technical, operational, and collaborative domains, enabling not only the detection of harmful content but also the protection of information integrity in communication infrastructures. Integrating these mechanisms into national and organizational resilience strategies is essential to safeguard public discourse and critical infrastructure. These tools exemplify the intersection of cybersecurity and misinformation response, providing both reactive and proactive mechanisms to detect, disrupt, and deter false information in the digital landscape. Key cybersecurity tools to counter misinformation include:

\begin{itemize}
    \item \textbf{Digital Services Act (DSA)} – An EU legislative framework that mandates large online platforms to identify, assess, and mitigate systemic misinformation risks, with enforceable accountability measures\footnote{\label{fn:DSA}https://commission.europa.eu/strategy-and-policy/priorities-2019-2024/europe-fit-digital-age/digital-services-act}.
    \item \textbf{Cyabra} – An AI-powered platform that collaboratively detects fake accounts, bot networks, and coordinated disinformation campaigns in real time\footnote{https://cyabra.com/} to enhance situational awareness and coordinated responses.
    \item \textbf{Fighting Disinformation Online} – A database of open-source web tools like factcheck.org\footnote {\label{fn:factchecker}https://www.factcheck.org/}, that uses detection algorithms to identify hashtags, links, accounts, and media being amplified by likely bots in a coordinated fashion, helping researchers and journalists track emerging disinformation campaigns\footnote{https://www.rand.org/research/projects/truth-decay/fighting-disinformation/search.html}.
    \item \textbf{SimSearchNet++ (Meta/Facebook)} – An advanced AI model that enhances image matching to detect variations and manipulations in images, such as cropping or blurring, which are common in misinformation campaigns. It also uses OCR to group and analyze text-based images\footnote{https://ai.meta.com/blog/heres-how-were-using-ai-to-help-detect-misinformation/}.
\end{itemize}

Such multifaceted strategies of leveraging technology and data underscore a proactive and integrated approach by governments to mitigate the impacts of misinformation. Through education, policy enforcement, transparency, collaboration, and technological innovation, governments are better equipped to safeguard public safety and enhance the resilience of critical infrastructure against the challenges posed by misinformation.

\section{Leveraging social infrastructure for community resilience}
To effectively counter misinformation within critical infrastructure sectors, social infrastructure -- including the extended network of professionals involved in maintaining the cyber, physical, and social facets of these systems -- can be leveraged. This broad community encompasses various stakeholders like utility workers, IT professionals, emergency responders, and local government officials, all pivotal in driving resilience efforts across different sectors. 
It is imperative that mechanisms are put in place to combat misinformation effectively to ensure that during times of crisis, truthful and helpful information prevails, aiding in efficient and safe disaster responses. There are many social infrastructure and community engagements that can be employed in this regard.

\subsection{Integrating human factors into resilience planning}

Resilience within the cyber-physical-social-climate nexus is heavily influenced by human factors, particularly those related to misinformation, communication failures, and lack of preparedness. Poor information-sharing, policy misalignment, and inadequate risk awareness can significantly undermine the ability of critical infrastructure systems to withstand and recover from disruptions. 

Failures in governance, policy misalignment, and institutional gaps have significantly impacted the resilience of critical infrastructures in past crises. One example is the 2021 ransomware attack on the Colonial Pipeline in the United States, which disrupted fuel supply across the East Coast. The attack exposed weaknesses in cybersecurity policies for privately operated infrastructure, where existing guidelines were merely advisory rather than mandatory, resulting in inconsistent security measures across sectors\footnote{https://www.cisa.gov/news-events/news/attack-colonial-pipeline-what-weve-learned-what-weve-done-over-past-two-years}. This incident highlights the necessity for enforceable cybersecurity policies that standardize best practices across all critical infrastructure systems.

The integration of cyber, physical, and social dimensions is essential to addressing these vulnerabilities. Decision-makers must adopt strategies that enhance preparedness, improve coordination, and boost awareness to manage the interconnected risks posed by cyber threats, physical disruptions, and climate risks \cite{Papamichael2024performing, Zhuo23}. For instance, research on mapping social networks for climate adaptation in Shoalhaven, Australia, has revealed the importance of key nodes in disseminating information effectively \cite{Cunningham_etal_2014}. Such findings emphasize the necessity of robust coordination among local actors to mitigate misinformation risks, infrastructure failures, and climate-related threats \cite{satizabal2022power}. 
Climate-induced risks also demonstrate the need to integrate human factors into resilience planning. For example, extreme flooding events have repeatedly overwhelmed urban drainage systems due to outdated flood management policies. Many cities lack the necessary infrastructure upgrades suggested by evolving climate models, leaving them vulnerable to extensive damage and economic loss. The absence of proactive climate resilience planning underscores the risks of inadequate adaptation measures. 
Similarly, the 2019/2020 Australian bushfire crisis showcased how poor coordination and communication between various government agencies exacerbated disaster response efforts. Delays in information-sharing and unclear jurisdictional responsibilities led to inefficient resource allocation and hindered evacuation efforts~\cite{inquiry2020final}. These cases illustrate how human vulnerabilities — whether due to policy gaps, communication failures, or lack of preparedness — can result in cascading failures that jeopardize public safety and infrastructure integrity. Transparent communication about the benefits of cybersecurity measures and climate adaptation strategies is essential to building public trust and overcoming resistance. Organizations can foster confidence by openly sharing how these measures mitigate risks and protect critical infrastructure, particularly in the face of climate-induced disruptions \cite{UNSG2023, UNICEF_BakuPrinciples}.  

Addressing human vulnerabilities in resilience planning requires a \textit{multi-faceted approach} that strengthens policies, enhances communication, and fosters systemic preparedness. Implementing \textit{enforceable regulations} is essential to standardizing cybersecurity measures and climate adaptation strategies, ensuring that misinformation does not hinder critical decision-making. Additionally, \textit{improving coordination and communication} between government agencies, infrastructure operators, and local stakeholders can help prevent disaster mismanagement and mitigate misinformation-driven delays. Strengthening interagency collaboration and establishing clear communication channels ensures that accurate information reaches the public and decision-makers in a timely manner.

Beyond policy and coordination, fostering systemic resilience is crucial to equipping governments, infrastructure operators, and communities with the necessary \textit{risk mitigation frameworks, training programs, and crisis response mechanisms}. Proactive adaptation to emerging threats, combined with improved public education and institutional preparedness, enhances infrastructure resilience and community decision-making, along with organizational-level integrated security planning, where IT teams and senior management collaborate, can significantly reduce financial and operational impacts\footnote{https://www.rutherfordsearch.com/blog/2021/09}. Furthermore, businesses that incorporate climate risks into their cybersecurity frameworks demonstrate a commitment to resilience, thereby strengthening stakeholder confidence and setting a precedent for sustainable operations\footnote{https://www.earth-scan.com/blog/climate-risk}. By integrating these strategies, stakeholders can \textit{minimize vulnerabilities, counter misinformation, and build a more resilient system} capable of withstanding and recovering from evolving risks.

\subsection{Strengthening social behavior and collective action for resilience}
Governments and institutions are instrumental in building trust and encouraging participation through inclusive decision-making processes. By promoting public awareness campaigns and facilitating partnerships between sectors, these bodies help to reinforce societal norms that promote cooperation and shared responsibility. This collective action at all levels — individual, community, corporate, and governmental — ensures that resilience strategies are well-designed and effectively implemented. For example, community workshops on renewable energy or waste reduction have been shown to enhance collective action and strengthen societal norms around sustainability \cite{djinlev2024collective}. 

Community-driven solutions and local leadership are crucial in disseminating accurate information and educating the public. Local leaders serve as trusted sources of information, making them key players in countering misinformation and enhancing community resilience. For example, in Namibia, agriculture sector networks have effectively utilized stakeholder collaboration to disseminate accurate climate information, showcasing the power of local initiatives in enhancing resilience against misinformation \cite{ofoegbu2021collaboration}. Similarly, community workshops focusing on sector-specific issues like renewable energy or waste management can amplify societal norms around sustainability and resilience, further enhancing community engagement. 
The integration of community engagement and social infrastructure into resilience strategies is indispensable. By harnessing the collective power of communities and leveraging robust educational frameworks, societies can enhance their ability to combat misinformation effectively. This approach strengthens community resilience and supports the broader efforts to safeguard critical infrastructure from the myriad threats posed by misinformation and other emerging challenges.

Public education initiatives, such as cybersecurity training programs and climate preparedness workshops, are pivotal in equipping individuals and organizations to respond to evolving threats. These programs help professionals understand how interconnected risks, such as compromized digital infrastructure or manipulated climate data, can undermine mitigation efforts. Workplaces and public institutions are particularly critical in embedding these practices into daily routines. For example, regular incident response drills and cybersecurity awareness campaigns have been implemented in the ISO 27001 framework, ensuring that organizations remain prepared for cyber threats while maintaining operational continuity during climate-related disruptions \cite{ISOIAF2024ClimateChange}. By fostering a culture of awareness and resilience through these measures, organizations can effectively safeguard both their digital assets and critical infrastructure from emerging challenges. In addition, by integrating cybersecurity considerations into climate policies and encouraging collaboration between government agencies, businesses, and communities, organizations can foster a more resilient ecosystem that is resistant to misinformation. 

Finally, individuals should be encouraged to build their own social identity (i.e., perceive oneself as belong to social groups), collective efficacy (i.e., people’s shared beliefs about their group’s capability to accomplish collective tasks), and positive deviance (i.e., a behavior-change approach that deepens the successful actions of existing community members) while working towards these bigger social connections \cite{Cosentino2024Community}. 
Collectively, these government, community and individuals initiatives can strengthen social behaviors that will make individuals within those communities more resistant against misinformation.

\subsection{Addressing policy and institutional gaps for enhanced resilience}
To enhance the resilience of the cyber-physical-social-climate nexus effectively, a comprehensive and coordinated approach is essential to address existing policy and institutional gaps. This involves modernizing regulations to reflect the evolving threat landscape, incorporating clear enforcement mechanisms for operators of critical infrastructure, and integrating long-term climate resilience measures into infrastructure planning. Policies need to be comprehensive, enforceable, and adaptable, ensuring new projects are designed with climate risk assessments and appropriate adaptation strategies \cite{AustCyberStra2023, Criticalinfra2023}. Furthermore, enhancing cross-sector coordination is crucial to break down silos and foster collaboration among various sectors to align policies across cyber, physical, and social domains enhances system-wide resilience \cite{Zhu2021Risk}.

Establishing robust accountability and compliance mechanisms is also vital to ensure that resilience measures are more than just recommendations — they must be mandatory and effectively enforced \cite{UNSG2023,AustCyberStra2023}. By implementing stricter monitoring and compliance protocols, regulatory agencies can guarantee that resilience practices are adhered to, significantly reducing the risks of system failures. Collectively, these strategies ensure that cyber, physical, and social systems are not only prepared to handle current threats but are also equipped to deal with future challenges. Addressing these policy and institutional gaps allows governments and stakeholders to build a more resilient infrastructure framework, ready to prevent and withstand cascading failures that could arise from both anticipated and unforeseen challenges~\cite{ Critical5_2024,Papamichael2024performing}. 

% \subsection{Improving decision-making}
Improving decision-making within the cyber-physical-social-climate nexus necessitates a sophisticated and nuanced approach to managing the growing interdependencies between these systems, particularly as the frequency of climate-related disruptions increasingly challenges them. Decision-makers must shift away from traditional static risk assessments towards dynamic, data-driven strategies that utilize advanced decision-support tools and predictive analytics. Such tools enable real-time risk assessments that account for cyber vulnerabilities, climate risks, and infrastructural interdependencies. For example, AI-driven tools allow scenario-based modeling that helps visualize potential cascading failures, offering decision-makers the ability to test and refine response strategies proactively \cite{Papamichael2024performing,Criticalinfra2023}.

Furthermore, breaking down organizational silos through enhanced cross-sector collaboration is essential for effective decision-making. Integrating efforts across cybersecurity, physical infrastructure management, and climate adaptation through public-private partnerships and interdisciplinary task forces ensures a cohesive approach to risk assessment and management \cite{Zhu2021Risk}. Training programs and scenario-based exercises are crucial in mitigating cognitive biases and enhancing decision-makers' ability to handle complex, multi-sector disruptions. Additionally, adopting flexible governance models and integrating decision-making with real-time data significantly enhance the responsiveness to emerging threats, ensuring that infrastructure systems are robust and resilient against both current and future challenges~\cite{Papamichael2024performing,Criticalinfra2023}.

By closing institutional and policy gaps, while embedding resilience in both governance frameworks and public consciousness, stakeholders can ensure long-term adaptability. Strengthening public awareness and integrating cybersecurity and climate preparedness into routine practices are essential components of a proactive resilience strategy.

\section{Technological solutions and innovations to fight against misinformation}

In the ongoing battle against misinformation, technological solutions and innovations play a crucial role. By leveraging advancements in AI, machine learning (ML), and cybersecurity, organizations and governments can significantly enhance their ability to identify, analyze, and counteract misinformation. These technologies are not only essential for maintaining the integrity of information but also for protecting digital platforms from the pernicious effects of misinformation.

\subsection{AI and ML in countering misinformation}
AI and ML are leading the charge in technological interventions against misinformation, utilizing their capability to process vast amounts of data with unparalleled speed and accuracy. These AI and ML tools aid in detecting patterns of misinformation, tracking narratives, and supporting informed decision-making, although they require human oversight and continuous retraining \cite{brundage2020trustworthy}. These tools enable the identification of fake news articles, and manipulated images and videos. 

A notable application within this domain is natural language processing (NLP), which empowers AI systems to scrutinize news articles, social media posts, and other digital content for misleading cues or factually incorrect statements by analyzing human language \cite{rojas2024hierarchical}. 
An example is the use of AI by researchers in Australia and the UK to combat climate misinformation. They are developing models that detect and debunk false climate claims on social media, using data from sources known for spreading misinformation, such as denial blogs, biased opinion sites, and politicized media outlets. These models are trained for real-time fact-checking, providing a proactive tool in the fight against misinformation \cite{AIModel2023, CARDS2023}. 
Another key innovation is the Augmented Computer Assisted Recognition of Denial and Scepticism (CARDS) model, a sophisticated hierarchical machine learning tool. It specifically targets contrarian climate claims on Twitter, analyzing large datasets to identify misinformation patterns and categorize claims for automated responses. These AI technologies are adept at tracking the sources and spread of misinformation, which equips analysts with detailed insights necessary for targeted interventions~\cite{CARDS2023}.

In a related advancement, a techno-cognitive approach combines topic modeling with network analysis to detect specific contrarian claims about climate hazards. This method illuminates the associations between these claims and corporate funding, shedding light on the strategic dissemination of misinformation \cite{Technocognitive2023}. The adaptability of machine learning models is also crucial as they continuously learn from new data and evolve to counteract the changing tactics of those spreading misinformation. This ongoing adaptation is vital for maintaining an effective defence against the dynamic nature of misinformation threats.

\subsection{Cybersecurity measures for information integrity}

Cybersecurity measures play a pivotal role in safeguarding information integrity across digital platforms, particularly vital in an era where misinformation frequently exploits vulnerabilities in information systems. Establishing robust cybersecurity protocols is essential for protecting data integrity, preventing unauthorized access, and ensuring that information remains untampered with and reliable across dissemination channels. Key to these efforts are encryption technologies, which secure data both at rest and in transit, preventing unauthorized interception and alteration. Additionally, robust authentication processes are crucial, as they prevent unauthorized access to systems where sensitive information is stored, thereby reducing the risk of internal leaks or the spread of misinformation.

Internationally, the UN is leading initiatives such as the proposed \textit{Code of Conduct}, which aims to protect information integrity on digital platforms by addressing false narratives, coordinated inauthentic behavior, and harmful content while safeguarding freedom of expression \cite{UNCode2023}. Similarly, Research ICT Africa's \textit{Information Integrity Programme} delves into the impacts of digital misinformation and explores regulatory solutions to ensure transparency, accountability, and trustworthiness in digital environments. This programme highlights the essential role of cybersecurity in maintaining credible information ecosystems\footnote{https://researchictafrica.net/programme/information-integrity/}. 
The UN Secretary-General issued a policy brief on information integrity, which underscores the need for coordinated international action to address threats to information integrity, including the implementation of cybersecurity measures to guard against online hate speech, misinformation, and harmful content. These measures are critical in enhancing the reliability of digital communications \cite{UNSG2023}.

Moreover, cybersecurity is fundamental in preventing data manipulation and securing digital platforms from coordinated attacks, ensuring that automated tools such as AI models used for fact-checking operate effectively without interference \cite{brundage2020trustworthy}. To support these defences, cybersecurity teams must employ active monitoring systems capable of detecting and responding to threats in real time. These systems help identify unusual activities that could indicate an information breach or attempts to inject misinformation. Once a potential threat is detected, immediate countermeasures are deployed to isolate and neutralize the threat, thereby minimizing potential damage and maintaining the integrity of information systems.

While emerging cybersecurity frameworks (e.g., DSA, SimSearchNet++)\footnote{https://commission.europa.eu/strategy-and-policy/priorities-2019-2024/europe-fit-digital-age/digital-services-act}\textsuperscript{,}\footnote{https://ai.meta.com/blog/heres-how-were-using-ai-to-help-detect-misinformation/} offer promising avenues for combating misinformation, practical challenges persist in their global deployment. These include enforcement limitations, varying platform compliance, and technological gaps in detecting sophisticated disinformation such as deepfakes. Their efficacy varies based on institutional readiness, platform cooperation, and technological sophistication. Policy enforcement, interoperability, and public adoption remain significant barriers.

\section{Lessons learned and best practice recommendations}

Managing human error, particularly misinformation susceptibility, in the cyber-physical-social-climate nexus requires a multifaceted approach. The potential for widespread disruption is significant, and therefore addressing the challenges posed by misinformation in critical areas requires a concerted effort to enhance public awareness, promote accurate information dissemination, and implement robust countermeasures against misinformation. By ensuring that truthful and helpful information prevails, especially in scenarios involving critical infrastructure, stakeholders can better safeguard these essential systems against the disruptive impacts of misinformation, thereby enhancing overall resilience and security.

\subsection{Lessons learned}

The following lessons highlight the interplay between human error, systemic interdependencies, and cascading failures, offering insights into the challenges faced in managing misinformation effectively.

\subsubsection*{Human error and susceptibility as a major factor}
Misinformation thrives on cognitive biases, limited digital literacy, and emotional responses, making individuals particularly susceptible during crises. The rapid dissemination of false narratives often leads to misguided decision-making, delayed response efforts, and reduced trust in official sources. The two presented case studies illustrate how false claims about arson and renewable energy failures contributed to widespread confusion, demonstrating that misinformation can exploit human cognitive vulnerabilities, even when factual information is readily available.

\subsubsection*{Interdependency between human and systemic vulnerabilities}
Misinformation does not operate in isolation; it interacts with infrastructure, governance, and digital platforms, leading to widespread consequences. As seen with the Australian bushfires and European heatwaves, misleading claims did not just alter public opinion but disrupted emergency communications, strained resources, and fueled political polarization. This underscores the need for holistic countermeasures that integrate human behavioral understanding with systemic resilience strategies.

\subsubsection*{System interdependencies and cascading failures}
False information can trigger a chain reaction of failures across interconnected systems, intensifying the scale of disruption. Misinformation in the cyber domain affects social perceptions, which in turn influence policy decisions and physical infrastructure resilience. The failure to curb disinformation about energy grid stability during the heatwave contributed to panic consumption, further destabilizing an already strained system. Similarly, misinformation about fire origins in Australia redirected resources away from critical climate adaptation efforts, delaying long-term resilience-building strategies. The highlights that misinformation exacerbates crises by distorting decision-making, undermining policy, and eroding trust. Addressing it requires understanding human biases, systemic interdependencies, and cascading failures. These insights are crucial for developing targeted interventions to enhance resilience and prevent future disruptions.

\subsection{Recommendations}

Building on the lessons learned, the following recommendations aim to address misinformation challenges within the cyber-physical-social-climate nexus. Each recommendation is aligned with a key lesson and includes actionable best practices to mitigate misinformation risks and enhance systemic resilience.

\subsubsection*{Addressing human susceptibility to misinformation by enhancing digital literacy and public awareness}
Human error and cognitive biases play a significant role in the spread of misinformation, particularly in high-stress situations where individuals rely on heuristics rather than critical analysis \cite{Zhuo24}. To counteract this, targeted interventions should equip individuals with the skills and tools necessary to critically assess information and reduce susceptibility to false narratives.

\begin{itemize}
    \item \textbf{Digital and media literacy upskilling:} Introduce educational programs that teach individuals how to critically evaluate online information and recognize disinformation tactics \cite{UNVerified2020}.
    \item \textbf{Behavioral interventions:} Implement nudges, warnings, and credibility indicators in digital platforms to alert users about potentially misleading content~\cite{beiser2023climate}.
    \item \textbf{Community-based misinformation monitoring:} Empower local communities and trusted organizations to detect and counter misinformation in real time \cite{GCoM2024}.
    \item \textbf{Crisis communication training:} Equip emergency responders and policymakers with tools to recognize, debunk, and respond effectively to misinformation in real-time scenarios \cite{Zhuo24}.
\end{itemize}

\subsubsection*{Strengthening systemic resilience against misinformation by integrating misinformation resilience into crisis management and policy frameworks}

Misinformation interacts with infrastructure, governance, and digital platforms, leading to widespread consequences, including emergency response failures and political polarization \cite{Trijsburg2024Disinformation}. To mitigate these risks, resilience strategies must be embedded within crisis management and policy frameworks, ensuring a coordinated and proactive approach to counter false narratives.

\begin{itemize}
    \item \textbf{Early warning systems for misinformation:} Develop AI-driven misinformation monitoring tools to identify and flag disinformation in real time~\cite{schreiber2021defalsif}.
    \item \textbf{Multi-stakeholder coordination:} Foster collaboration between government agencies, technology companies, scientific institutions, and media to ensure rapid response to misinformation \cite{UNESCO2024,nato2024}.
    \item \textbf{Transparent communication strategies:} Ensure that official sources provide timely, transparent, and consistent updates to preempt false narratives \cite{EC2024}.
    \item \textbf{Fact-checking and rapid response units:} Establish dedicated teams within emergency agencies to detect, verify, and counter misinformation during disasters~[\ref{fn:factchecker}]~\cite{CAAD2024}.
\end{itemize}

Building on the systemic vulnerabilities discussed throughout this chapter, the NATO SPS Programme plays a critical role in fostering coordinated, transnational responses to climate-related misinformation. As NATO continues to refine its response to information threats\footnote{\url{https://www.nato.int/cps/en/natohq/topics_219728.htm}}, some relevant areas for strategic development and collaboration are identified:

\begin{itemize}
    \item \textbf{Cross-national misinformation monitoring:} Support interoperable tools for real-time tracking, sentiment analysis, and early warning of misinformation across NATO member states.
    \item \textbf{Crisis communication protocols:} Promote standardized protocols for verifying and disseminating official information during cyber-physical-climate crises.
    \item \textbf{Cyber hygiene integration:} Fund training and exercises to embed cyber hygiene practices, such as like multi-factor authentication, access controls, and security audits, across civil and military systems.
    \item \textbf{Misinformation intelligence consortium:} Establish a NATO-led consortium to coordinate disinformation threat analysis and share countermeasures across sectors.
    \item \textbf{Content provenance pilots:} Invest in blockchain and watermarking pilots to verify the authenticity of official digital content during high-risk events.
\end{itemize}

By investing in these strategic lines of effort, NATO SPS would not only help counter the destabilizing effects of misinformation but also reinforce transnational trust, operational cohesion, and public awareness—key ingredients for effective climate adaptation and security resilience across the Alliance.

\subsubsection*{Mitigating cascading failures from system interdependencies by developing proactive misinformation management strategies}

Misinformation can cause cascading failures across interconnected systems, amplifying disruptions in cyber, physical, social, and climate infrastructures \cite{jamalzadeh2022protecting}. To prevent misinformation from escalating into systemic crises, proactive strategies must be embedded into misinformation management efforts, ensuring early intervention and infrastructure resilience.

\begin{itemize}
    \item \textbf{Pre-bunking over debunking:} Implement pre-bunking strategies that educate the public on common misinformation tactics before crises occur \cite{beiser2023climate}.
    \item \textbf{Resilient infrastructure planning:} Integrate misinformation resilience into the design of smart grids, emergency communication systems, and disaster management plans \cite{Critical5_2024}.
    \item \textbf{Accountability measures for disinformation actors:} Implement regulatory frameworks to hold individuals and organizations accountable for deliberate misinformation dissemination\textsuperscript{[\ref{fn:DSA}]}.
    \item \textbf{Empirical testing of counter-misinformation measures:} Conduct real-world simulations and behavioral studies to refine misinformation counter-strategies based on empirical evidence \cite{CARDS2023}.
\end{itemize}

To summarize, countering misinformation demands proactive strategies across individual, institutional, and systemic levels. Strengthening digital literacy, embedding resilience in policies, and reinforcing infrastructure will help mitigate risks. Implementing countermeasures ensures robust crisis management, infrastructure security, and climate adaptation efforts.

\section{Challenges and future directions}

Misinformation poses a severe threat to critical infrastructure by undermining public trust, delaying emergency responses, and exacerbating vulnerabilities during climate-related hazards and natural disasters. Addressing this challenge requires proactive measures such as improving communication accuracy, debunking false narratives swiftly, and building public resilience through education on disaster preparedness and climate science. 
By effectively countering misinformation, governments and organizations can enhance infrastructure resilience and ensure more robust responses to future crises, safeguarding lives and essential services from preventable failures~\cite{WEF2024}. The following key challenges highlight the primary obstacles in managing misinformation effectively.

\begin{itemize}
    \item  \textbf{Political and institutional barriers:} Climate misinformation is often driven by political and economic interests, leading to hesitancy in implementing countermeasures due to censorship concerns and geopolitical tensions.
    \item  \textbf{Technological limitations:} AI tools struggle to keep pace with evolving disinformation tactics like deepfakes and bot networks, while social media’s decentralized nature allows false narratives to spread before fact-checking can intervene.
    \item  \textbf{Social and behavioral challenges:} Cognitive biases, ideological beliefs, and low digital literacy make misinformation difficult to counter, as false narratives often reinforce pre-existing opinions despite factual corrections.
    \item  \textbf{Interdependencies and cascading failures:} Misinformation interacts with cyber, physical, and social systems, triggering widespread disruptions such as panic-driven overconsumption, policy resistance, and delayed crisis responses, necessitating a coordinated approach.
\end{itemize}

Considering the current landscape and future challenges, effective misinformation management must go beyond high-level goals and be grounded in domain-specific interventions. Table~\ref{tab:misinformation_summary} summarizes these actionable strategies across the cyber, physical, social, and climate dimensions of the CPS-Climate nexus. For each domain, it outlines the primary impacts of misinformation, targeted countermeasures, and relevant data sources for monitoring progress. These strategies respond directly to the challenges identified above: political and institutional barriers are addressed through transparency and governance, technological limitations through AI-driven detection and platform accountability, social and behavioral challenges via literacy, trusted messengers, and behavioral nudges, and cascading failures through systemic integration of misinformation resilience. By operationalizing these responses, stakeholders can move from reactive misinformation control to proactive resilience-building across critical systems.

\section{Conclusion}

The cyber-physical-social-climate nexus intricately connects diverse domains that underpin the functionality and resilience of modern society. Each domain — cyber, physical, social, and climate — plays a distinct role, yet their interdependencies mean that disruptions in one can cascade across the others. For instance, digital components of physical infrastructure may rely on data connectivity or environmental monitoring, both of which are vulnerable to climate-driven disruptions. These systems are further compromized by misinformation, which complicates public understanding, weakens institutional trust, and impairs critical decision-making.

This chapter systematically addressed these challenges through the four primary objectives introduced in Section 1. First, the impact of climate misinformation on interconnected systems was examined in detail in Sections 2 and 3. Second, Section 3.1 analyzed how communication failures, whether due to misinformation or technological faults, can escalate crises. Third, countermeasures including governance strategies, technological interventions, and cybersecurity frameworks were explored in Sections 4 and 6. Finally, Sections 5 and 7 outlined actionable resilience strategies that leverage social infrastructure and institutional reforms to improve systemic adaptability and public trust.

\begin{table}[h]
\centering
\renewcommand{\arraystretch}{1.3}
\begin{tabular}{|p{1.5cm}|p{3.1cm}|p{3.5cm}|p{3.5cm}|}
\hline
\textbf{Dimension} & \textbf{Impact of misinformation} & \textbf{Possible way forward} & \textbf{Example Metrics or Data Sources} \\ \hline

\textbf{Cyber} & Spread of false narratives through social media, disruption of emergency alerts, fake donation scams. & Strengthen misinformation monitoring with AI-driven detection, enforce platform accountability, and improve digital literacy initiatives. & EU DSA\textsuperscript{~\ref{fn:DSA}}; number of verified fact-checks flagged\textsuperscript{~\ref{fn:factchecker}}. \\ \hline

\textbf{Physical} & Misallocation of emergency resources, public non-compliance with evacuation orders, infrastructure strain. & Implement early warning misinformation response systems, integrate misinformation awareness in emergency training, and enhance crisis communication strategies. & Response lag time \cite{GCoM2024, downtoearth2025falsealerts}; ND-GAIN Infrastructure Indicators\textsuperscript{9}; public adherence to evacuation orders \cite{GCoM2024}. \\ \hline

\textbf{Social} & Erosion of trust in authorities, increased climate scepticism, political polarization, resistance to climate adaptation. & Promote transparency in government communication, increase public engagement in policy decisions, and strengthen media literacy education. & Edelman Trust Barometer \cite{edelman2023}; social media sentiment analysis \cite{islam2021covid,pertwee2022epidemic,UNVerified2020}; participation rates in climate policy forums \cite{dablanorris2023public}. \\ \hline

\textbf{Climate} & Delay in climate adaptation policies, undermining of climate science, increased vulnerability to future wildfires. & Integrate misinformation resilience into climate policies, enhance fact-checking collaborations with scientific institutions, and conduct pre-bunking campaigns to counter climate misinformation. & ND-GAIN Readiness Score\textsuperscript{\ref{fn:ndgain}}; IPCC policy adoption lags \cite{ipcc2023}. \\ \hline
\end{tabular}
\caption{Misinformation impact and possible way forward}
\label{tab:misinformation_summary}
\end{table}

Table \ref{tab:misinformation_summary} provides a comprehensive overview of the challenges presented by misinformation within the cyber-physical-social-climate nexus, as discussed in this chapter. This offers insights into how these challenges can be effectively managed and mitigated through coordinated efforts across multiple domains. It examined the impact of climate misinformation on the cyber-physical-social nexus, highlighting how communication failures and misinformation increase vulnerabilities in critical infrastructure. It discussed combating climate misinformation through governance, community initiatives, and technology. Finally, it presented actionable strategies to enhance critical infrastructure resilience and societal functions against climate misinformation.

This chapter emphasized the need for a holistic, cross-domain approach to protect societal functions such as energy supply, healthcare delivery, and public safety, against the growing threats posed by climate risks and information disruption. Resilience in this context means not only securing individual systems but also safeguarding the interconnections between them. Achieving this requires a comprehensive strategy that integrates proactive engagements, cross-sector collaboration, and informed decision-making. To support this, the chapter explores vulnerabilities rooted in human behavior within the cyber-physical-social-climate nexus, examining how actions, decisions and behaviors can undermine the resilience and adaptability of these interconnected systems.

\begin{acknowledgement}
The research supporting this chapter was funded by CSIRO's Critical Infrastructure Protection and Resilience (CIPR) mission, which aims to strengthen resilience and foster informed dialogue around the critical infrastructure that underpins economic, environmental, and societal systems.
\end{acknowledgement}

\bibliographystyle{ieeetr}
\bibliography{sample}

\begin{thebibliography}{10}

\bibitem{haque2021realizing}
M.~A. Haque, S.~Shetty, K.~Gold, and B.~Krishnappa, ``Realizing cyber-physical systems resilience frameworks and security practices,'' {\em Security in Cyber-Physical Systems: Foundations and Applications}, pp.~1--37, 2021.

\bibitem{ipcc2023}
{Intergovernmental Panel on Climate Change (IPCC)}, ``{Sixth Assessment Report (AR6)}.'' \url{https://www.ipcc.ch/assessment-report/ar6/}, 2021--2023.

\bibitem{lse2024}
{LSE Grantham Institute}, ``What are climate misinformation and disinformation and what is their impact?.'' \url{https://www.lse.ac.uk}, 2024.

\bibitem{nature2023}
T.~Spampatti, U.~J.~J. Hahnel, E.~Trutnevyte, and T.~Brosch, ``Psychological inoculation strategies to fight climate disinformation,'' {\em Nature Human Behaviour}, vol.~7, p.~1136–1148, 2023.

\bibitem{grobler2024building}
M.~Grobler and T.~Aamir, ``Building cognitive resilience for enhanced cyber governance,'' pp.~52--72, 2024.

\bibitem{cook2019understanding}
J.~Cook, ``Understanding and countering misinformation about climate change,'' in {\em Handbook of Research on Deception, Fake News, and Misinformation Online} (I.~Chiluwa and S.~Samoilenko, eds.), pp.~281--306, IGI Global, 2019.

\bibitem{UNVerified2020}
{United Nations Department for Global Communications}, ``The {UN} globally launches {`Verified'} campaign to combat misinformation about {COVID-19}.'' \url{https://www.undp.org/kyrgyzstan/press-releases/un-globally-launches-verified-campaign-combat-misinformation-about-covid-19}, 2020.

\bibitem{Zhuo24}
S.~Zhuo, R.~Biddle, J.~D. Recomendable, G.~Russello, and D.~M. Lottridge, ``Eyes on the phish(er): Towards understanding users' email processing pattern and mental models in phishing detection,'' in {\em Proceedings of the 2024 European Symposium on Usable Security, EuroUSEC 2024, KarlstadSweden, 30 September 2024- 1 October 2024} (F.~Karegar and A.~Farooq, eds.), pp.~15--29, {ACM}, 2024.

\bibitem{inquiry2020final}
{NSW Bushfire Inquiry}, ``Final report of the {NSW Bushfire} inquiry.'' \url{https://www.dpc.nsw.gov.au/assets/dpc-nsw-gov-au/publications/NSW-Bushfire-Inquiry-1630/Final-Report-of-the-NSW-Bushfire-Inquiry.pdf}, 2020.

\bibitem{islam2021covid}
M.~S. Islam, A.-H.~M. Kamal, A.~Kabir, D.~L. Southern, S.~H. Khan, S.~M. Hasan, T.~Sarkar, S.~Sharmin, S.~Das, T.~Roy, {\em et~al.}, ``Covid-19 vaccine rumors and conspiracy theories: The need for cognitive inoculation against misinformation to improve vaccine adherence,'' {\em PloS one}, vol.~16, no.~5, p.~e0251605, 2021.

\bibitem{pertwee2022epidemic}
E.~Pertwee, C.~Simas, and H.~J. Larson, ``An epidemic of uncertainty: {R}umors, conspiracy theories and vaccine hesitancy,'' {\em Nature medicine}, vol.~28, no.~3, pp.~456--459, 2022.

\bibitem{edelman2023climate}
{Edelman}, ``{2023 Edelman Trust Barometer Special Report: Trust and Climate}.'' \url{https://www.edelman.com/sites/g/files/aatuss191/files/2023-11/2023%20Edelman%20Trust%20Barometer%20Special%20Report%20Trust%20and%20Climate.pdf}, 2023.

\bibitem{schreiber2021defalsif}
D.~Schreiber, C.~Picus, D.~Fischinger, and M.~Boyer, ``The defalsif-ai project: protecting critical infrastructures against disinformation and fake news,'' {\em e\&i Elektrotechnik und Informationstechnik}, vol.~138, no.~7, pp.~480--484, 2021.

\bibitem{hameleers2023disinfo}
M.~Hameleers, ``Disinformation as a context-bound phenomenon: {T}oward a conceptual clarification integrating actors, intentions and techniques of creation and dissemination,'' {\em Communication Theory}, vol.~33, no.~1, pp.~1--10, 2023.

\bibitem{maertens2020combatting}
R.~Maertens, F.~Anseel, and S.~van~der Linden, ``Combatting climate change misinformation: Evidence for longevity of inoculation and consensus messaging effects,'' {\em Journal of Environmental Psychology}, vol.~70, p.~101455, 2020.

\bibitem{GCoM2024}
{Global Covenant of Mayors for Climate \& Energy}, ``Addressing rampant climate disinformation,'' report, Global Covenant of Mayors for Climate \& Energy, October 2024.

\bibitem{CISC2024RR}
{Cyber and Infrastructure Security Centre}, ``Critical infrastructure annual risk review: Second edition,'' report, {Department of Home Affairs, Australian Government}, November 2024.
\newblock Available \href{https://www.cisc.gov.au/resources-subsite/Documents/critical-infrastructure-annual-risk-review-2024.pdf}, Accessed on February 07, 2025.

\bibitem{owen2019nuclear}
L.~H. Owen, ``Nuclear disasters, information vacuums: How a lack of data in {Fukushima} led to the spread of fake health news.'' \url{https://www.niemanlab.org/2019/07/nuclear-disasters-information-vacuums-how-a-lack-of-data-in-fukushima-led-to-the-spread-of-fake-health-news/}, 2019.

\bibitem{WEF2024}
{World Economic Forum}, ``The global risks report 2024,'' report, World Economic Forum, Geneva, Switzerland, January 2024.

\bibitem{ecker2022psychological}
U.~K. Ecker, S.~Lewandowsky, J.~Cook, P.~Schmid, L.~K. Fazio, N.~Brashier, P.~Kendeou, E.~K. Vraga, and M.~A. Amazeen, ``The psychological drivers of misinformation belief and its resistance to correction,'' {\em Nature Reviews Psychology}, vol.~1, no.~1, pp.~13--29, 2022.

\bibitem{kozyreva2023incorporating}
A.~Kozyreva, L.~Smillie, and S.~Lewandowsky, ``Incorporating psychological science into policy making,'' {\em European psychologist}, 2023.

\bibitem{dablanorris2023public}
E.~Dabla-Norris, S.~Khalid, G.~Magistretti, and A.~Sollaci, ``Public support for climate change mitigation policies: A cross country survey,'' {\em IMF Working Papers}, vol.~2023, no.~223, pp.~1--48, 2023.

\bibitem{OECD2022}
{OECD}, ``Building trust and reinforcing democracy: Preparing the ground for government action,'' tech. rep., Nov. 2022.

\bibitem{oilprice2024misinformation}
OilPrice.com, ``The impact of misinformation on wind energy development,'' 2024.
\newblock Accessed: 2025-05-08.

\bibitem{longview2014wind}
{The Long View Group}, ``Summary of the literature: Wind farms and health,'' 2014.
\newblock Available \href{https://eisdocs.dsdip.qld.gov.au/Coopers%20Gap%20Wind%20Farm/Final%20EIS/Final%20EIS%20-%20Appendices/appendix-h-summary-of-the-literature-wind-farms-and-health.pdf}{here}, Accessed: 2025-05-08.

\bibitem{roozenbeek2023countering}
J.~Roozenbeek, E.~Culloty, and J.~Suiter, ``Countering misinformation,'' {\em European Psychologist}, 2023.

\bibitem{Trijsburg2024Disinformation}
I.~Trijsburg, H.~Sullivan, E.~Park, M.~Bonotti, P.~Costello, Z.~Nwokora, D.~Pejic, M.~Peucker, and W.~Ridge, ``Disinformation in the city: Response playbook,'' tech. rep., The University of Melbourne, 2024.

\bibitem{jamalzadeh2022protecting}
S.~Jamalzadeh, K.~Barker, A.~D. González, and S.~Radhakrishnan, ``Protecting infrastructure performance from disinformation attacks,'' {\em Scientific Reports}, vol.~12, no.~1, p.~16832, 2022.

\bibitem{Broda2024Misinformation}
E.~Broda and J.~Str{\"o}mb{\"a}ck, ``Misinformation, disinformation, and fake news: lessons from an interdisciplinary, systematic literature review,'' {\em Annals of the International Communication Association}, vol.~48, no.~2, pp.~139--166, 2024.

\bibitem{Hilberts2024TheIO}
S.~Hilberts, S.~Evers, M.~Govers, and E.~Petelos, ``The impact of misinformation on social media in the context of natural disasters,'' {\em The European Journal of Public Health}, vol.~34, 2024.

\bibitem{downtoearth2025falsealerts}
{Down To Earth}, ``False alerts like the one sent during the {Greater Los Angeles} wildfires can undermine trust and provoke anxiety.'' \url{https://www.downtoearth.org.in/climate-change/false-alerts-like-the-one-sent-during-the-greater-los-angeles-wildfires-can-undermine-trust-and-provoke-anxiety}, 2025.

\bibitem{bushfire_disinformation}
T.~Graham, ``Bushfires, bots and arson claims: {A}ustralia flung in the global disinformation spotlight,'' {\em The Conversation}, Jan 2020.

\bibitem{edelman2023}
{Edelman}, ``{2023 Edelman Trust Barometer}.'' \url{https://www.edelman.com/trust/2023/trust-barometer}, 2023.

\bibitem{Heffernan_2024}
A.~Heffernan, ``Disinformation in {Africa},'' digital policy hub — working paper, Centre for International Governance Innovation (CIGI), Winter 2024.

\bibitem{liu2025climate}
J.~C.-E. Liu and C.-F. Lee, ``Climate and energy misinformation in {Taiwan},'' {\em Frontiers in Communication}, vol.~9, p.~1531126, 2025.

\bibitem{strauss2024reporting}
N.~Strauss, J.~Painter, J.~Ettinger, M.-N. Doutreix, A.~Wonneberger, and P.~Walton, ``{Reporting on the 2019 European heatwaves and climate change: Journalists' attitudes, motivations and role perceptions},'' in {\em Journalism and Reporting Synergistic Effects of Climate Change}, pp.~226--249, Routledge, 2024.

\bibitem{casero2023european}
A.~Casero-Ripoll{\'e}s, J.~Tu{\~n}{\'o}n, and L.~Bouza-Garc{\'\i}a, ``The {European} approach to online disinformation: {Geopolitical} and regulatory dissonance,'' {\em Humanities and Social Sciences Communications}, vol.~10, no.~1, pp.~1--10, 2023.

\bibitem{kaur2024reag}
S.~T. Kaur, T.~Aamir, S.~Trewin, and M.~Grobler, ``{REAG (Resilience Expert Advisory Group)} pre-workshop survey quantitative analysis findings,'' 2024.

\bibitem{Critical5_2024}
{Critical 5}, ``Adapting to evolving threats: A summary of critical 5 approaches to critical infrastructure security and resilience,'' Government Report PS9-35/2024E-PDF, Public Safety Canada, Ottawa, Canada, June 2024.
\newblock In partnership with the governments of Australia, Canada, New Zealand, United Kingdom, and United States.

\bibitem{UNESCO2024}
{UNESCO}, ``Global initiative for information integrity on climate change.'' \url{https://www.unesco.org/en/information-integrity-climate-change}.

\bibitem{EC2024}
{European Commission}, ``Climate disinformation.'' climate.ec.europa.eu.

\bibitem{AustraliaC2021}
{Department of Agriculture, Water and the Environment}, ``National climate resilience and adaptation strategy 2021 – 2025: {Positioning Australia} to better anticipate, manage and adapt to our changing climate.'' \url{https://www.agriculture.gov.au/sites/default/files/documents/national-climate-resilience-and-adaptation-strategy.pdf}, 2021.

\bibitem{unesco2024information}
{UNESCO}, ``Global initiative for information integrity on climate change.'' \url{https://www.unesco.org/en/information-integrity-climate-change}, 2024.

\bibitem{G202024}
{United Nations}, ``{G20 Leaders' Summit Brazil: UN and UNESCO} launch global initiative on information integrity.'' \url{https://www.un.org/en/climatechange/page/g20-leaders%E2%80%99-summit-brazil-un-and-unesco-launch-global-initiative-information}, 2024.

\bibitem{WMO2024}
{World Meteorological Organization (WMO)}, ``{WMO} joins global initiative for information integrity to fight climate disinformation.'' WMO, Nov 2024.

\bibitem{CAADOpenLetter2024}
{Climate Action Against Disinformation (CAAD) coalition and climate experts}, ``Open letter: Governments should act now to curb climate disinformation.'' \url{https://caad.info/wp-content/uploads/2024/11/Open-Letter_-Governments-Should-Act-Now-to-Curb-Climate-Disinformation-1.pdf}.

\bibitem{CAAD2024}
{Climate Action Against Disinformation (CAAD)}, ``Climate action against disinformation’s response to the senate inquiry on greenwashing.'' \url{https://www.aph.gov.au/DocumentStore.ashx?id=0ed5ee92-5eee-44f5-a288-724aad574f55&subId=744106}, 2025.

\bibitem{Papamichael2024performing}
M.~Papamichael, C.~Dimopoulos, and G.~Boustras, ``Performing risk assessment for critical infrastructure protection: an investigation of transnational challenges and human decision-making considerations,'' {\em Sustainable and Resilient Infrastructure}, pp.~1--19, 2024.

\bibitem{Zhuo23}
S.~Zhuo, R.~Biddle, Y.~S. Koh, D.~M. Lottridge, and G.~Russello, ``{SoK: H}uman-centered phishing susceptibility,'' {\em {ACM} Trans. Priv. Secur.}, vol.~26, no.~3, pp.~24:1--24:27, 2023.

\bibitem{Cunningham_etal_2014}
R.~Cunningham, C.~Cvitanovic, T.~Measham, B.~Jacobs, A.-M. Dowd, and B.~Harman, ``A preliminary assessment into the utility of social networks for engaging local communities in climate adaptation policy,'' tech. rep., Institute for Sustainable Futures and CSIRO, Nov 2014.
\newblock Working Paper Prepared for: NSW Office of Environment and Heritage.

\bibitem{satizabal2022power}
P.~Satiz{\'a}bal, I.~Cornes, M.~d. L.~M. Zurita, and B.~R. Cook, ``The power of connection: Navigating the constraints of community engagement for disaster risk reduction,'' {\em International Journal of Disaster Risk Reduction}, vol.~68, p.~102699, 2022.

\bibitem{UNSG2023}
{United Nations}, ``Code of conduct for information integrity on digital platforms,'' tech. rep., Dec 2023.

\bibitem{UNICEF_BakuPrinciples}
{UNICEF}, ``Baku guiding principles on human development for climate resilience.'' \url{https://www.unicef.org/media/165126/file}.

\bibitem{djinlev2024collective}
V.~Djinlev and B.~J. Pearce, ``Collective action lessons for the energy transition: {L}earning from social movements of the past,'' {\em Sustainability Science}, vol.~19, pp.~847--863, 2024.

\bibitem{ofoegbu2021collaboration}
C.~Ofoegbu and M.~New, ``Collaboration relations in climate information production and dissemination to subsistence farmers in {Namibia},'' {\em Environmental Management}, vol.~67, no.~1, pp.~133--145, 2021.

\bibitem{ISOIAF2024ClimateChange}
{ISO 9001 Auditing Practices Group}, ``Guidance on: Auditing climate change issues in {ISO 9001},'' edition 1, International Organization for Standardization (ISO) and International Accreditation Forum (IAF), Mar. 2024.

\bibitem{Cosentino2024Community}
M.~Cosentino, R.~Gal-Oz, and D.~L. Safer, ``Community-based resilience: The influence of collective efficacy and positive deviance on climate change-related mental health,'' in {\em Storytelling to Accelerate Climate Solutions}, pp.~319--338, Springer International Publishing Cham, 2024.

\bibitem{AustCyberStra2023}
{Department of Home Affairs}, ``2023--2030 {Australian} cyber security strategy,'' government report, Australian Government, Canberra, Australia, November 2023.

\bibitem{Criticalinfra2023}
{Australian Government Department of Home Affairs}, ``Critical infrastructure annual risk review: First edition,'' 2023.
\newblock Available \url{https://www.cisc.gov.au/resources-subsite/Documents/critical-infrastructure-annual-risk-review-first-edition-2023.pdf}.

\bibitem{Zhu2021Risk}
T.~Zhu, S.~Haugen, and Y.~Liu, ``Risk information in decision-making: Definitions, requirements and various functions,'' {\em Journal of Loss Prevention in the Process Industries}, vol.~72, p.~104572, 2021.

\bibitem{brundage2020trustworthy}
M.~Brundage, S.~Avin, J.~Clark, H.~Toner, P.~Eckersley, B.~Garfinkel, A.~Dafoe, P.~Scharre, T.~Zeitzoff, B.~Filar, H.~Anderson, H.~Roff, G.~Allen, J.~Steinhardt, C.~Flynn, R.~Kagan, G.~Leung, T.~Maharaj, M.~Trager, {\em et~al.}, ``Toward trustworthy ai development: Mechanisms for supporting verifiable claims,'' {\em arXiv preprint arXiv:2004.07213}, 2020.

\bibitem{rojas2024hierarchical}
C.~Rojas, F.~Algra-Maschio, M.~Andrejevic, T.~Coan, J.~Cook, and Y.-F. Li, ``Hierarchical machine learning models can identify stimuli of climate change misinformation on social media,'' {\em Communications Earth \& Environment}, vol.~5, no.~1, p.~436, 2024.

\bibitem{AIModel2023}
T.~G. Coan, C.~Boussalis, J.~Cook, and M.~O. Nanko, ``Computer-assisted classification of contrarian claims about climate change,'' {\em Scientific reports}, vol.~11, no.~1, p.~22320, 2021.

\bibitem{CARDS2023}
C.~Rojas, F.~Algra-Maschio, M.~Andrejevic, T.~Coan, J.~Cook, and Y.-F. Li, ``Augmented cards: A machine learning approach to identifying triggers of climate change misinformation on twitter,'' {\em arXiv preprint arXiv:2404.15673}, 2024.

\bibitem{Technocognitive2023}
F.~Zanartu, J.~Cook, M.~Wagner, and J.~Garc{\'\i}a, ``A technocognitive approach to detecting fallacies in climate misinformation,'' {\em Scientific Reports}, vol.~14, no.~1, p.~27647, 2024.

\bibitem{UNCode2023}
{United Nations}, ``Information integrity on digital platforms: Our common agenda policy brief 8,'' tech. rep., June 2023.

\bibitem{beiser2023climate}
L.~F. Beiser-McGrath, T.~Bernauer, R.~Gampfer, M.~R. Sisco, and S.~van~der Linden, ``Climate action with, by and for the people,'' {\em Nature Human Behaviour}, vol.~7, pp.~1655--1667, 2023.

\bibitem{nato2024}
{NATO CPSICC Workshop}, ``Cpsicc nexus workshop 2024: Addressing climate change impacts on cyber-physical-social systems.'' \url{https://cpsiccnexusworkshop2024.org}, 2024.

\end{thebibliography}

\end{document}